\documentclass[final,3p,times,twocolumn]{elsarticle}
\usepackage{graphicx}
\usepackage{epsfig}
\usepackage{amssymb}
\usepackage{amsmath}

\def\local#1{\langle #1 \rangle}
\def\secref#1{Sec.~\ref{#1}}
\def\figref#1{Fig.~\ref{#1}}

\def\eff{{eff}}
\def\td{{\rm d}}
\def\nonum{\nonumber}
\def\vn{{\bf v}_n}

\def\ns{n_s}
\def\vs{{\bf v}_s}
\def\js{{\bf j}_s}
\def\jn{{\bf j}_n}
\def\jl{{\bf j}_l}
\def\uv{{\bf u}_v}
\def\zz{\hat{\bf z}}
\def\molf{{\bf P}}
\def\emas{m_e}     
\def\Ex{E_{\|}}
\def\Ey{E_\bot}
\def\Re{{{\rm Re}\ }} 
\def\Im{{{\rm Im}\ }}

\def\clampy{`clamped'}

\def\curl{\nabla\times}
\def\del{\partial}

\def\Aout{{\bf A}_{out}}

\usepackage{color}
\graphicspath{{./images/}}  
\journal{Physica C}
\begin{document}
\begin{frontmatter}
  \title{Transverse acousto-electric effect in superconductors}
  \author[1]{P.~Lipavsk\'y}
  \ead{lipavsky@karlov.mff.cuni.cz \sep http://fu.mff.cuni.cz/theory/lipavsky/}
  \author[2]{J. Kol{\'a}{\v c}ek}
  \ead{kolacek@fzu.cz}
  \author[3]{P.-J.~Lin}
  \ead{fareh.lin@gmail.com}
  \address[1]{Faculty of Mathematics and Physics,
    Charles University, Ke Karlovu 3, 12116 Prague 2, Czech Republic}
  \address[2]{Institute of Physics, Academy of Sciences,
    Cukrovarnick{\'a} 10, 162 00 Prague 6, Czech Republic}
  \address[3]{Research Department, Universal Analytics Inc., Airdrie, AB T4B 2A4, 
  Canada}

  \begin{abstract}

    We formulate a theory based on the time-dependent Ginzburg Landau
    (TDGL) theory and Newtonian vortex dynamics to study the transverse
    acousto-electric response of a type-II superconductor with Abrikosov
    vortex lattice. When exposed to a transverse acoustic wave, Cooper pairs 
    emerge from the the moving atomic lattice and moving electrons. 
    As in the Tolman-Stewart effect in a normal metal, an electromagnetic 
    field is radiated from the superconductor.  We adapt the equilibrium-based 
    TDGL theory to this non-equilibrium system by using a floating condensation 
    kernel. Due to the interaction between normal and superconducting 
    components, the radiated electric field as a function of magnetic field 
    attains a maximum value occurring below the upper critical magnetic field. 
    This local increase in electric field has weak temperature dependence and 
    is suppressed by the presence of impurities in the superconductor.
  \end{abstract}
  \begin{keyword}
    superconductivity \sep acousto-electric effect \sep Abrikosov vortex 
    \sep Tolman-Stewart effect
    \PACS 74.20.De \sep 74.25.Ld \sep 72.50.+b
  \end{keyword}
\end{frontmatter}

\section{Introduction}
\label{Introduction}

Superconductivity appears at low temperatures when materials are rigid
and fragile so that for the majority of experiments made in cryostat
it is not necessary to consider any motion of the crystal. The
standard time-dependent Ginzburg-Landau theory (TDGL), which contains
the assumption of local equilibrium, and is formulated in the
laboratory frame, is a powerful tool to study phenomena in the
vicinity of the superconducting-normal phase transition.  To study
gyroscopes \cite{Gaw93}, gravitational wave antennae \cite{GFNNST13}
and the interaction of the superconducting condensate with strong
sound waves
\cite{HL96,Sonin96,Gutlyanskii98,FFAGJ05,Fil06,JH07,HJ09,HCMPK00,Gutlyanskii02,Gutlyanskii03,AC08},
where the atomic lattice is in motion, an extension to TDGL is needed to
accommodate the dynamical system.

In the presence of a transverse acoustic wave, the condensate
does not experience friction with the crystal and its imperfections as
Cooper pairs do not scatter on the underlying crystal or its
imperfections. The moving lattice, however, acts on the
condensate by three nondissipative mechanisms:
(i) Induction:
The motion of ions creates an electric current which affects
the electrons by magnetic induction.
(ii) Entrainment: A Cooper pair of zero momentum in a moving crystal
has nonzero velocity with respect to the crystal.
This effect is particularly strong in dirty superconductors,
where the mass of the Cooper pair is strongly renormalized, $m^*\gg 2\emas$.
In the reference frame locally moving with the lattice, entrainment
together with the fictitious force compose the inertial force causing
the Tolman-Stewart effect.
(iii) Deformation potentials: Deformations of the crystal lead to
local changes of the chemical potential and material parameters which
control superconductivity, for example its critical temperature.

Because of the induction the supercurrent tends to oppose the ionic
current, but the compensation is not always complete. For example, in
a steadily-rotating (but stationary) superconductor, currents near the
surface are only partly compensated and the residual current produces
a magnetic field known as the London moment \cite{L50}. In
non-stationary (oscillating) systems the compensation is even less
effective. In particular, under the influence of the ultrasound wave,
the inertial motion of normal electrons as well as superconducting
electrons leads to nonzero bulk currents via the Tolman-Stewart
effect. The oscillating current radiates electromagnetic waves
\cite{FFAGJ05,Fil06} so that the superconductor exhibits nonzero
acousto-electric effect.

Theoretical analysis of the acousto-electric field in superconductors
has been performed assuming the fully superconducting state with no
normal current \cite{Fil06}; the normal electrons are \clampy ~to
the lattice \cite{Sonin96,Gutlyanskii98,Gutlyanskii02,Gutlyanskii03}.
Using this assumption, the TDGL theory of Verkin and Kulik \cite{VK72}
(originally developed for the case of steady rotation when normal
currents are absent) can be used to study the acousto-electric field.

Experimental studies on the acousto-electric field of hole-type metal
(niobium) and electron-type borocarbide
(Y$_{0.95}$Tb$_{0.05}$Ni$_2$B$_2$C) show a 10\% increase of the
radiated electric field as the material transforms into the
superconducting state (Fig.~2 in Ref.~\cite{Fil06}).  This small
change shows that, at least in the vicinity of the transition
temperature, the normal and the superconducting currents are
comparable, which is incompatible with the assumption of stationary
(\clampy) normal electrons.  Our aim is to develop a theory which
accounts for coexisting normal and superconducting currents.

Here we study theoretically the effect of a transverse sound wave on a
superconducting system near the superconducting-normal
phase-transition line, $B_{c2}$; the transverse wave propagates along
the $z$ axis, and oscillates in the $x$ direction.  In the transverse
wave the material experiences shear stress only, without compression.
According to the experimental studies by Fil {\em et al} \cite{Fil06},
the changes of potential and material parameters caused by shear
deformations can be disregarded.  To accommodate this non-equilibrium
system, we use a modified version of TDGL theory based on a
microscopic derivation \cite{LBK13}.

In \secref{GL} we formulate the set of TDGL equations
for the dynamical system with oscillating atomic lattice driven by 
an external transverse acoustic wave. Motion of normal electrons is allowed 
and this current treated in a self-consistent way, instead of assuming the 
normal electrons move together \clampy ~to the ions as in \cite{Fil06,Sonin96}.
As a result, the normal current is driven by inertial forces as in the 
normal-state Tolman-Stewart effect. Details of our derivation based on the 
Boltzmann equation are given in \ref{A1}. In \secref{Vortex} using Newtonian 
dynamics, we analyze the acousto-electric effect in the mixed state.  Vortex 
dynamics in a steady state is deduced from the force balance on vortices. 
Magnus, pinning and transverse forces are considered, along with friction 
forces from the atomic lattice and from normal electrons 
\cite{KK76,Iordanskii66}.  The effective forces acting on the 
superconducting electrons are identified from the extended TDGL 
equation in \secref{GL}. Next we discuss the skin effect and the
matching of the internal field and the radiated electromagnetic wave 
at the surface. \secref{Vortex} councludes with the resulting complete 
set of equations. \secref{Numpred} contains numerical computations of 
the radiated electric field, using material parameters provided in 
\cite{Fil06}.

%#################################
\section{TDGL theory}
\label{GL}
%#################################

To study the electrons in an oscillating atomic lattice, it is
advantageous to choose the moving background as the reference frame as
Cooper pairs emerge from the moving electrons.  Previous study
\cite{LBK13} shows that to apply standard TDGL theory to a dynamical
system, it is optimal to choose a condensation kernel floating with
the background.  Here we omit the details and write down the set of
equations known as the floating-kernel time-dependent Ginzburg-Landau
(FK-TDGL) equations.

Relaxation of the Ginzburg-Landau (GL) order parameter $\psi$ in the 
dynamical system is described by the FK-TDGL equation
\begin{eqnarray}
  \frac{1}{2m^*}(-i\hbar\nabla-e^*{\bf A}-\molf )^2\psi
  &-&\alpha\psi+\beta|\psi|^2\psi \nonum\\
  &=&\Gamma\bigg(\frac{\del}{\del t}-i\frac{2}{\hbar}\mu\bigg)\psi
  \label{FK1}
\end{eqnarray}
with the molecular field
\begin{equation}
  \molf =\chi^* m^*{\dot{\bf u}}+m^*\vn.
  \label{molfield}
\end{equation}
The first term of the molecular field is due to the entrainment effect
caused by motion of the ionic lattice with velocity $\dot{\bf u}$; the
mass of a Cooper pair is $m^*=m_0/(1+\chi^*)$, where $m_0=2\emas $ is
twice the electron mass $\emas $.
The corresponding superconducting current is
\begin{equation}
  \js={e^*\over m^*}\Re \big[
    \bar\psi
    (-i\hbar\nabla-e^*{\bf A}-\molf )\psi \big].
  \label{FKcur}
\end{equation}

The velocity $\vs$ of the condensate can be defined using
$\js=e^*|\psi|^2\vs = e\ns\vs$.
Because of the presence of the
second term $m^*\vn$ in $\molf$, the operator
$(1/m^*)(\!-i\hbar\nabla\!-\!e^*{\bf A}\!-\!\molf )$ gives the
velocity of Cooper pairs with respect to normal electrons $\vn$.
The current generated by the moving ions is
\begin{equation}
  \jl=-en\dot{\bf u},
  \label{ioncur}
\end{equation}
where ${\bf u}$ is the ion displacement caused by the
transverse sound wave.

In our treatment we relax the requirement of Sonin \cite{Sonin96} that
the electrons move with the same velocity $\dot{\bf u}$ as ions.
Instead we assume that as with the Tolman-Stewart effect in normal
conducting metals \cite{Davis88}, the normal electrons lag behind ions
and move with velocity $\vn$. The normal current $\jn = e n \vn$ can be
obtained from the Boltzmann equations, shown in \ref{A1}; the electric
current is
\begin{equation}
  \jn + \jl=\frac{\sigma_{n}}{e}
  \bigg(
       {\bf F}'-\frac{\tau}{1+i\tau\omega}\frac{e}{m}
       {\bf B}\times{\bf F}'
       \bigg)-\nu\dot{\bf u}.
       \label{Ohm}
\end{equation}
The effective driving force
\begin{equation}
  {\bf F}'=-e{\del{\bf A}\over\del t}-\nabla\mu+
  e\dot{\bf u}\times{\bf B}-\emas \ddot{\bf u}
  \label{efforceText}
\end{equation}
includes the effective electric field (first and second terms),
a part of the Lorentz force ${\bf F}_{L}=
e({\bf v}\rq + \dot{\bf u})\times{\bf B}$, where ${\bf v}\rq$ is
electron velocity relative to the lattice, and the inertial force.
These terms can be understood in the reference frame
moving with the lattice, where the third term enters the
electric field via a Lorentz transformation.
The relaxation time $\tau$ comes from the normal conductivity
$\sigma_{n}$, from \eqref{sigma}.
The last term in relation \eqref{Ohm} results from the diffusion of
the transverse momentum \cite{Fil01}; this is similar to the mechanism
causing the shear viscosity.
Detail derivation of \eqref{Ohm}, analogous to Ohm's law, from the
Boltzmann equation can be found in \ref{A1}.

From the continuity equation $\nabla \cdot {\bf j}=0$ we can obtain for 
the chemical potential 
\begin{equation}
  \nabla^2\mu={e\over\sigma_{n}}\nabla\cdot\js+	
  e\nabla\cdot(\dot{\bf u}\times{\bf B}),	
  \label{potential}
\end{equation}
which is simplified by the transversality condition ${\bf q}\cdot{\bf u}=0$ 
for wave vector ${\bf q}$. The total force has zero divergence, so 
$\nabla\cdot\ddot{\bf u}=0$.
We consider a system with homogeneous conductivity, $\nabla\sigma_{n}=0$.

The vector potential ${\bf A}$ can be obtained from the Maxwell equation
\begin{equation}
  \nabla^2{\bf A}=-
  \mu_0 ( \js + \jn + \jl );
  \label{Maxwell}
\end{equation}
we use the Coulomb gauge $\nabla\cdot{\bf A}=0$.
To obtain the radiated electromagnetic wave, we must evaluate
skin vector potential and match internal and external fields.
In \secref{Me}, we will show that the skin effect is
negligible if the wavelength of radiation is much larger
than the skin depth.

We have a three-component system consisting of normal electrons,
condensate, and electromagnetic field.  Equations
\eqref{FK1}, \eqref{FKcur}, \eqref{Ohm}, \eqref{potential} and
\eqref{Maxwell} form a complete set of equations of motion.
We are interested below in \secref{Vortex} a case that the transverse
sound wave interacts with a superconductor in the mixed state.
Here we compare our theory with the TDGL theory of Verkin and Kulik~\cite{Fil01,VK72,Sonin96} referred as VK-TDGL.

To make the comparison,
we rewrite our equations in terms of the relative velocities with respect 
to the atomic lattice, that is, the relative velocity of normal electrons 
as $\vn'=\vn-\dot{\bf u}$ and the relative velocity of the condensate as 
$\vs'=\vs-\dot{\bf u}$.

In this notation, the molecular field \eqref{molfield} is
\begin{equation}
  \molf =m_0{\dot{\bf u}}+m^*\vn'.
  \label{molfieldVerkin}
\end{equation}
The first term is the fictitious force obtained by Verkin and Kulik \cite{VK72}.
The second term which is absent in \cite{VK72} is a correction due to
non-zero velocity of normal electrons with respect to the ionic lattice.
From Ohm's law \eqref{Ohm}, \eqref{sigma} and \eqref{sharevisc},
we can see that the relative velocity is proportional to mean
free path $\ell$. In the dirty limit ( $\ell\ll\xi_0$) $\vn'\to {\bf 0}$, 
hence the second term in \eqref{molfieldVerkin} can be ignored; our theory 
then reduces to VK-TDGL.

%#######################################
\section{Vortex dynamics}
\label{Vortex}
%#######################################

Near the normal and superconducting phase transition, vortex motion
is well described by TDGL theory. By solving the TDGL equation with
the assumption of rigid Abrikosov vortex lattice, the TDGL equation
can be represented in the form of force balance of Newtonian equations.
Here we consider a superconductor which occupies $z<0$ with
rigid Abrikosov vortices; each vortex has a fluxon $\Phi_0$ along
the $z$-axis; the magnetic induction is ${\bf B}=(0,0,B)$, $B>B_{c1}$;
thus the interspacing between vortices is $a\sim\sqrt{\Phi_0/B}$.
The transverse acoustic wave propagates along the $z$-axis and oscillates
in the $x$ direction; the atomic lattice deformation can be evaluated as
the real part of the complex function
${\bf u}\equiv \exp(i\omega t)\cos(qz) (u,0,0)$.

This physical system contains variables at microscopic scale, such as
parameters describing motion of electrons, and variables at
mesoscopic scale, such as the wavelength of the acoustic wave.
The typical wavelength of the acoustic wave is $\sim 100$ ${\rm \mu}$m,
and that of the radiation is of the order of a metre,
while spacing between vortices is $\sim 100$ nm.

Since we are interested in phenomena at mesoscopic scale, we can
average a microscopic field $f$ locally to produce a mesoscopic field
$\local{f}$, by writing
\begin{equation}
 \local{f}(t,{\bf r}) := 	
 (B/\Phi_0) \int_{C_{\bf r}} \td x'\td y' f(t,{\bf r}')
\end{equation}
where the 2-D region $C_{\bf r}$ is the size and shape of an elementary cell,
but with centroid ${\bf r}$ rather than being aligned with the lattice.

The acoustic wave acts on the superconductor in a similar manner as
far-infrared (FIR) light; the condensate accelerates, so
$\local{\vs}\neq0$.
Following the idea by Sonin \cite{Sonin96},
we will use the theory of vortex motion derived and experimentally
tested for FIR response to study the interaction of
acoustic waves with a superconductor in the mixed state.

%--------------------------------
\subsection{Balance of forces on the condensate}
\label{Bfc}
%--------------------------------

Under the influence of the transverse acoustic wave, the effective force
driving Cooper pairs into motion can be identified as the time derivative
of the effective vector potential ${\bf A}_\eff ={\bf A}+\molf /e^*$ in \eqref{FK1}:
\begin{equation}
 \local{{\bf F}}=-\frac{\del}{\del t}	
 \local{e^*{\bf A}+\molf }.
 \label{efFsup}
\end{equation}
This averaged local force is balanced by
\begin{equation}
 \local{{\bf F}}=	
 -e^*\dot\uv\times\local{{\bf B}}+	
 m^*\frac{\del{\local{\vs}}}{\del t},
 \label{Josephson}
\end{equation}
where $\dot\uv$ is the velocity of Abrikosov vortices.
The first term comes from induction and by itself would comprise the
Josephson relation \cite{Josephson65}; the second term is essential in
dynamical systems when $\dot{\local{\vs}} \neq 0$. The full
\eqref{Josephson} is known as the inertial Josephson relation (IJR) which
can be obtained either from the standard TDGL theory
\cite{LLM12,Matlock12} or hydrodynamic theory \cite{AKK65}.
Gutlyanski{\v\i} uses an identical equation (Eqn.~(1) of
\cite{Gutlyanskii98}) which he calls the London equation because of
its inertial term.

%--------------------------------
\subsection{Balance of forces on vortices}
\label{Vd}
%--------------------------------
In a dynamical situation, a vortex experiences a number of forces
\cite{Kop01}.  The simplest equation of motion for a vortex is through
the compensation of the Lorentz force and the Bardeen-Stephen friction
force due to dissipative scattering of quasi-particles in the
vortex-core region.  In reality, impurity of a sample complicates the
dynamics of vortices; it leads to such things as vortex pinning on the
mesoscopic scale, and modified relaxation times of particles on the
microscopic scale.

Here we adopt a widely-used equation from \cite{Kop01}.
Considering all the forces
on a vortex of unit length in the oscillating lattice,
\begin{eqnarray}
 e\ns(\dot{ \uv}-\local{\vs})\times {\bf z}&=&	
 \eta_{lat}(\dot\uv-\dot{\bf u})	\nonumber\\	
 &+&\eta_{qp}(\dot\uv-\local{\vn})	\nonumber\\
 &+&\alpha_{L}(\uv-{\bf u})		\nonumber\\
 &+&\alpha_{KK}(\dot\uv-\dot{\bf u})\times{\bf z}	\nonumber\\
 &-&\alpha_{I}(\dot\uv-\local{\vn})\times{\bf z},
 \label{vortexmot}
\end{eqnarray}
where we have defined ${\bf z} = \local{{\bf B}_\eff }/\local{B_\eff }$.
The left side contains the hydrodynamic Lorentz and Magnus forces.
The Magnus force can be obtained from the TDGL equation with the
assumption of a rigid Abrikosov lattice. The Lorentz force comes from the
effective magnetic field $\local{{\bf B}_\eff }=\curl\local{{\bf A}_\eff }$.
In the linear response region $\dot\uv-\local{\vs}$ is small, therefore
the effect of the acoustic wave on the magnetic field can be neglected, so
$\local{{\bf B}_\eff } \approx \local{{\bf B}}$ and we may write
${\bf z} \approx\local{{\bf B}}/\local{B} = \zz$.

The forces on the right side of \eqref{vortexmot} arise from a more
detailed microscopic picture. The first two terms are frictional
forces of vortex with the ionic lattice, and with normal electrons.
The pinning force with the Labush parameter $\alpha_{L}$ is
proportional to the relative displacement of the vortex from a pinning
centre fixed to the ionic lattice.  The transverse force of Kopnin and
Kravtsov \cite{KK76}, due to scattering on impurities, has coefficient
$\alpha_{KK}$.  Finally, the interaction of the vortex with
quasiparticles has the transverse component of Iordanskii type
\cite{Iordanskii66} with coefficient $\alpha_{I}$.

Rewriting \eqref{vortexmot} in terms of relative velocities
$\dot{\uv'}$, $\vs'$ and $\vn'$,
we can separate the force imposed by the normal current, writing
\begin{eqnarray}
-e\ns\local{\vs'}\times {\bf z}&=&
\eta\dot\uv\rq{}
+\alpha_{L}\uv\rq{}
-\alpha_{M}\dot\uv\rq{}\times{\bf z}
\nonumber\\
&&-\eta_{qp} \local{\vn'}+\alpha_{I} \local{\vn'}\times{\bf z},
 \label{vortexmotrel}
\end{eqnarray}
where $\eta=\eta_{lat} + \eta_{qp}$ is the total coefficient of
friction, and $\alpha_{M}=e\ns-\alpha_{KK} + \alpha_{I}$ accounts for
both corrections to the Magnus force.

In the limit  $\vn' \to {\bf 0}$, corresponding to the
\clampy ~electron model, \eqref{vortexmotrel} coincides with
the equation for forces on the vortex lattice used by Fil {\em et al}
\cite{Fil06}.
This limit is justified for dirty materials where $\tau\omega\ll1$ in
Fil's measurement ($\tau\sim 10^{-13}$ s and frequency $55$ MHz).
The theory developed here without restriction to this limiting case
is valid for moderately pure materials and higher frequencies near
sub-gap frequencies.

%----------------------------------------
\subsection{Skin effect and the Maxwell equation}
\label{Me}
%----------------------------------------

The vector potential inside the superconductor contains a large
contribution ${\bf A}^0$ from the Abrikosov vortex lattice which satisfies
$\del_t\local{{\bf A}^0}=0$, and a time-dependent perturbation which
is the sum of the internal field with space dependence given by the
acoustic wave $\local{{\bf A}'}= {\bf A}' \exp(i\omega t)\cos(qz)$
and the skin field $\local{{\bf A}''}={\bf A}''\exp(i\omega t+z/\lambda_{sk})$
where $\lambda_{sk}$ is the skin depth.

We solve for the surface fields by writing
$\local{\vs'}(t,{\bf r}) = \vs' \exp(i\omega t) \cos(qz)$
(similarly for $\vn$ and other fields), shifting our notation so that
from here onwards; $\vs'$ refers to the field at the surface at $t=0$.

The Maxwell equation \eqref{Maxwell} gives
\begin{equation}
 q^2 {\bf A}'=
 \mu_0e\left(\ns\vs'+i\omega \ns{\bf u}+n\vn'\right).
 \label{Maxwellred}
\end{equation}
$\vs'$ depends only on the oscillating transport current, as
the strong static diamagnetic currents forming vortices
average to zero over a cell.

At the surface, the matching of the internal
electromagnetic wave 
${\bf A}'(t, {\bf r})+{\bf A}''(t, {\bf r})$ and
the outgoing radiation $\Aout \exp(i\omega(t-z/c))$ yields two conditions.
The first condition, obtained from the Maxwell equation
${\bf E}=-\del_t {\bf A}$, is
\begin{eqnarray}
\Aout&=&{\bf A}'+{\bf A}''.
\label{surfmatch1}
\end{eqnarray}
The second condition, obtained from ${\bf B}=\curl{\bf A}$, is
\begin{eqnarray}
-i\frac{\omega}{c}\Aout&=&\frac1{\lambda_{sk}}{\bf A}''.
 \label{surfmatch2}
\end{eqnarray}
In \eqref{surfmatch2} we used that the rotation of the field
is proportional to $\sin(qz)$ which vanishes at the surface.
Solving for the radiated field from \eqref{surfmatch1} and
\eqref{surfmatch2}, we find
\begin{equation}
\Aout=\frac1{1+i\frac{\omega}{c}\lambda_{sk}}
{\bf A}'.
 \label{matching}
\end{equation}
Since the wavelength of the radiation is $c/\omega\gg\lambda_{sk}$, we
can approximate the radiated field by the internal one, so $\Aout={\bf A}'$.

%----------------------------
\subsection{Equations for surface fields}
\label{Final}
%----------------------------
The ionic displacement ${\bf u}$ is known. The vortex
displacement $\uv'$, the condensate velocity $\vs'$,
the normal velocity $\vn'$, and the vector potential ${\bf A}'$
are required. Here we rewrite equations in a convenient
form.

The vortex displacement given by \eqref{vortexmotrel} at
frequency $\omega$ is
\begin{eqnarray}
	e \ns \zz\times\vs'&=&
		( i\omega\eta + \alpha_L )\uv'
		+ i\omega\alpha_{M}\zz\times\uv\rq{}\nonumber\\
		&&-\eta_{qp}\vn'-\alpha_{I}\zz\times\vn'.
\label{vortexmotred}
\end{eqnarray}

The condensate velocity is obtained from the IJR \eqref{Josephson} with
the force \eqref{efFsup}
\begin{equation}
	2e{\bf A}'=-2m\omega_{c}\zz\times(\uv'+{\bf u})
		-2i\omega \emas {\bf u}
		-m^*(\vs'+\vn'+i\omega{\bf u}),
 \label{Josephsonred}
\end{equation}
where $\omega_{c}=eB/m$ is the cyclotron frequency.

The normal velocity is obtained from Ohm's law \eqref{Ohm};
the electric field in the force \eqref{efforceText} is needed.
Using the periodicity of the Abrikosov vortex lattice, we obtained
$\local{\nabla\mu} = 0$ in \ref{Chemical}.
Together with ${\bf B}\cdot{\bf u}=0$, the normal velocity is
\begin{eqnarray}
m\vn'&=&-
\frac{i\tau\omega}{ 1+i\tau\omega}\bigg[
e{\bf A}'- \frac{\tau\omega_{c}}{ 1+i\tau\omega}
\,\zz\times e{\bf A}'+
\nonumber\\ &+&
\bigg(i\omega \emas +
m\omega_{c}\frac{\tau\omega_{c}}{ 1+i\tau\omega}+
\frac{e\nu}{\sigma_{n}}\bigg){\bf u}
\nonumber\\ &+&
\bigg( m\omega_{c}+
i\omega \emas
\frac{\tau\omega_{c}}{ 1+i\tau\omega}\bigg)
\,\zz\times{\bf u}\bigg].
\label{Ohmredfullred}
\end{eqnarray}

We have considered radiation in response to a transverse acoustic
wave; dependent on magnetic field, temperature and relaxation time,
the radiation can now be evaluated by solving \eqref{Maxwellred},
\eqref{vortexmotred}-\eqref{Ohmredfullred}.
With ${\bf B}$ and ${\bf q}$ along the $z$ axis,
all of ${\bf u}$, $\uv'$, $\vs'$, $\vn'$ and ${\bf A}'$ have zero
$z$-components. We have eight algebraic equations for $x$ and $y$
components of four unknown vectors.

Our model relaxes the assumption of VK-TDGL theory that normal
electrons are stationary with respect to ions, but neither treatment
takes into account effects of thermal fluctuations which are
particularly strong at the superconducting and normal phase
transition. The effect of thermal fluctuations can be included by the
introduction of Langevin forces $\zeta({\bf r}, t)$ in the left side
of the FK-TDGL equation \eqref{FK1}, which equation describes the
dynamics of the order parameter. In this paper, we will next restrict
our discussion to the case of conventional superconductors where the
macroscopic fluctuations are negligible.

%%%%%%%%%%%%%%%%%%%%%%%%%%%%%%%%
\section{Numerical predictions}
\label{Numpred}
%%%%%%%%%%%%%%%%%%%%%%%%%%%%%%%%
We study numerically the acousto-electric effect in the mixed state of
a superconductor and in the region beyond the superconducting-normal
phase transition line $B_{c2}$. In the normal state, our theory
reproduces the Tolman-Stewart effect.  The distance to the phase
transition line is defined as $\delta b=(B-B_{c2}(T))/B_{c2}(0)$.  The
normal state corresponds to $\delta b>0$, the superconducting state to
$\delta b<0$.  Residual-resistance ratio (RRR) measured in the normal
state is used to quantify the effect of imperfection of the atomic
crystal.  Here we focus on the case of niobium; necessary material
parameters are taken from the measurement by Fil {\em et al}
\cite{Fil06}.  Parameters regarding the forces on vortices are
specified in \ref{Foronvor}.

\begin{figure}
\hskip-4mm
\includegraphics[scale=1.05]{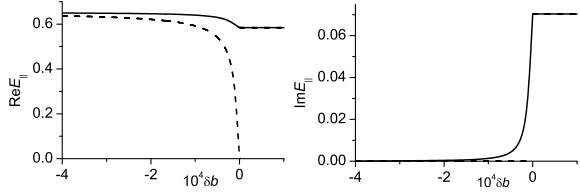}
\caption{Parallel acousto-electric coefficient as function of magnetic
field for superconductor with an RRR of $62$:
$\Ex$ as a function of $\delta b$ near the critical line.
The dashed line corresponds to the VK-TDGL model, and is not defined for positive $\delta b$.
The solid line shows the FK-TDGL result, and is continuous in $\delta b$.
\label{figEx12}}
\end{figure}

We first discuss the $\delta b$ dependence of the radiated electric
field $\Ex$, parallel to the atomic displacement due to the transverse
acoustic wave incident perpendicular to the surface.  The radiated
electric field is normalized by its magnitude at $B=0$ and $T=0$,
where all electrons are in the condensate.  Shown in \figref{figEx12}
is the radiation calculated from two different models, the FK-TDGL
theory and the VK-TDGL theory, for a sample with an RRR value of $62$
at $T=0.75T_c$; this corresponds to Fil's Fig.~3 of \cite{Fil06}.

The overall radiation at $T=0.75T_c$ is smaller then the radiation at
$T=0$ and $B=0$; the radiation increases when entering superconducting
state and saturates as $B \rightarrow 0$.  As expected, FK-TDGL theory
gives a non-zero $\Ex$ in the normal state, consistent with the Tolman-
Stewart value, but VK-TDGL theory which ignores the effect of normal
current gives $\Ex=0$.  Nevertheless, both theoretical curves show
marked changes near the phase transition, and the two curves coincide
at small magnetic field where superconductivity is robust.

Both of these models suggest that $\Im\Ex$ is negligible for very
negative $\delta b$. Near the phase transition, the FK-TDGL model
indicates a fundamental increase when approaching $\delta b=0$ and then
remains constant.  The VK-TDGL curve shows that $\Im\Ex$ remains
negligible throughout the whole superconducting region.

\begin{figure}
\hskip-4mm
\includegraphics[scale=1.05]{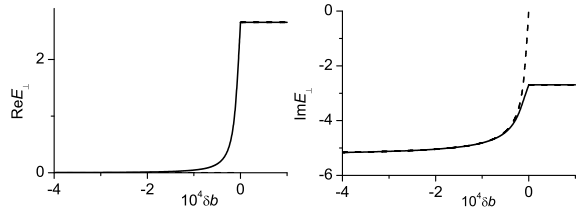}
\caption{Transverse acousto-electric coefficient as a function of the
$\delta b$ for RRR of $62$: $\Ey$, shown by the solid line,
is continuous in FK-TDGL theory while the dashed line shows a
`step' appearing in the VK-TDGL model.
\label{figEy12}}
\end{figure}

$\Re \Ey$ is dominated by the normal current and has similar behaviour
to $\Im \Ex$. In \figref{figEy12}, $\Re \Ey$ in the FK-TDGL model
increases with $\delta b$ before the abrupt change near $\delta b=0$,
while $\Re \Ey$ remains negligible according to the VK-TDGL model.
$\Im \Ey$ curves in the two models increase and coincide in the
superconducting state; they separate when approaching $\delta b=0$.
The VK-TDGL curve goes to zero at $\delta b=0$, while the FK-TDGL curve
shows a continuous change through transition into the normal
state.

\begin{figure}
\hskip-4mm
\includegraphics[scale=1.05]{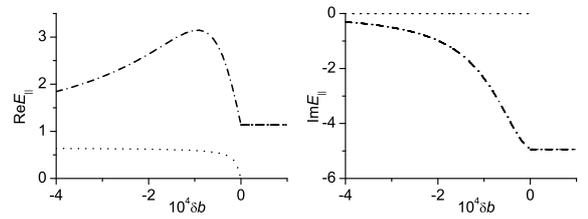}
\caption{Parallel acousto-electric coefficient as a function of
  magnetic field for a superconductor with RRR of $620$ at $T=0.75
  T_c$: The dot-dashed line shows the FK-TDGL result and the dotted
  line shows that of VK-TDGL; this convention is to facilitate
  comparison with the plots in \figref{figExT}.
}
\label{figEx34}
\end{figure}

According to the FK-TDGL model, imperfections of a superconductor
influence radiation near the phase transition. The FK-TDGL model suggests
that a cleaner superconductor with RRR of $620$ emits stronger
radiation near the phase transition, shown in \figref{figEx34}. The
maximum $\Re \Ex$ is around three times larger than its value in the
normal state, or its value in the purely-superconducting state at $T=0$
and $B=0$.
The VK-TDGL plot shows the radiated electric field increasing
with $\delta b$ as in the dirtier superconductor
shown in \figref{figEx12}; the effect of impurities is negligible.
The off-phase component $\Im \Ey$ in \figref{figEx34} is suppressed
in the superconducting state as in the dirtier superconductor discussed
previously (with a different sign).

\begin{figure}
\hskip-4mm
\includegraphics[scale=1.05]{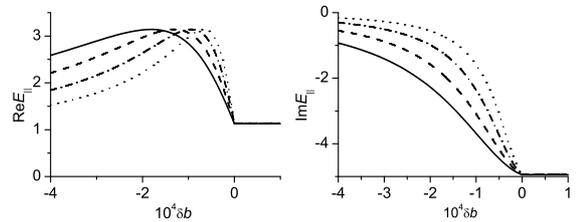}
\caption{$\Ex$ as a function of the $\delta b$ for RRR of $620$:
  $t=0.1$ (solid line); $t=0.5$ (dashed line); $t=0.75$ (dot-dashed
  line); $t=0.99$ (dotted line).
}
\label{figExT}
\end{figure}

The enhancement of the radiation due to interaction between
superconducting current and normal current is temperature dependent.
$\Ex(\delta b)$ of the FK-TDGL model plotted at various temperatures
is shown in \figref{figExT}.  The location of the maximum gradually
moves away from $\delta b=0$ as temperature decreases. While the peak
is widened as temperature decreases, the magnitude of the peak changes
little.

In contrast to the VK-TDGL model, FK-TDGL model accounts for the
interaction between superconducting current and normal current.  Our
model provides a continuous description for a superconducting system
in transition to normal state, and shows the Tolman-Stewart effect of
normal metal in the normal state.  When the superconducting system is
away from the phase transition, the normal current contributes less
and our model coincides with the VK-TDGL model. The VK-TDGL model is
justified also in a dirty superconductor because the normal electrons
scatter with impurities and thereby tend to move together with the
lattice. However, in a clean superconductor the normal current
contributes to the radiation;  our model shows that
radiation is enhanced due to the interaction between superconducting
current and normal current. This enhancement occurs in superconducting
state near the phase transition, and the field can reach three times that
of radiation emitted in the normal state for certain values of $\delta b$.

%%%%%%%%%%%%%%%%%%%%%%%%%%%%%%%%
\section{Summary}
%%%%%%%%%%%%%%%%%%%%%%%%%%%%%%%%
The acousto-electric effect has been shown in the vicinity of the
critical magnetic field to reveal the interference of the
superconducting and the normal response. To investigate this
interference, we have employed the time-dependent Ginzburg-Landau
theory, taking into account the effect of the normal current on the
formation of the condensate.  This formulation with normal current had
been derived earlier from the microscopic approach within the
framework of a floating nucleation kernel.

The Ginzburg-Landau theory with the inertial term of
Verkin and Kulik provides reliable predictions, save for within a very
narrow vicinity of the critical line between the normal and the
superconducting state. This deficiency is emphasized in cleaner samples.
The interference appearing in this narrow vicinity shows enhancement
which we expect to be observable, in particular in the case of Niobium
with a high RRR.

\section*{Acknowledgments}
This work was supported by M{\v S}MT COST projects LD15062 and LD14060.
P.-J. Lin acknowledges financial support from UA through SR-621-1207.
Authors are grateful to V. D. Fil and D. V. Fil for many discussions
on this subject and P. Matlock for critical reading.

%%%%%%%%%%%%%%%%%%%%%%
\appendix
%%%%%%%%%%%%%%%%%%%%%%
%-----------------------------------------------------------------
\section{Transverse acousto-electric effect in normal metal}
\label{A1}
%-----------------------------------------------------------------
In this appendix we derive the interaction of the transverse acoustic
wave with the normal metal. The wave propagating in the $z$ direction,
with wave vector ${\bf q} = (0,0,q)$, is described by the
amplitude of lattice deviation in the $x$-direction,
${\bf u} = (u,0,0)$, with
$u = ue^{i\omega t - i {\bf q}\cdot{\bf r} }$.

As the wave is transverse, the electric current generated by it
is also transverse, therefore $\nabla\cdot{\bf j}=0$. From the
equation of continuity follows that the charge density does not
change $\del_t\rho=-\nabla\cdot{\bf j}=0$ so that we can set
$\phi=0$. We note that this argument holds to the linear order in
${\bf u}$. At quadratic order, there is a small charge transfer along the
$z$ axis e.g. due to the Bernoulli effect; we neglect quadratic effects.

The generated transverse electric field is covered by the Maxwell
equation $-\nabla^2{\bf A} = \mu_0{\bf j}$.
Using $-\del_t{\bf A}={\bf E}$ yields
\begin{equation}
q^2{\bf E}=-i\omega\mu_0{\bf j}.
\label{MaxwellE}
\end{equation}

We need to evaluate the current as a function of the electric field
${\bf E}$ and the deviation ${\bf u}$. To this end we use the
Boltzmann equation in the relaxation time approximation
\begin{equation}
\frac{\del f}{\del t}+{\bf v}\cdot\nabla f-
\nabla\varepsilon\cdot\frac{\del f}{\del {\bf k}}
=-{1\over\tau}\delta f,
\label{Boltzmann}
\end{equation}
where $\delta f=f-{\bar f}$ is a deviation from local equilibrium.
The local equilibrium distribution ${\bar f}$ represents electrons
emitted from collisions with impurities and lattice vibrations. It
has the same local density as the actual distribution
\begin{equation}
2\int{\td{\bf k}\over (2\pi)^3}{\bar f}=
2\int{\td{\bf k}\over (2\pi)^3}f=n,
\label{relaxn}
\end{equation}
where the factor of two accounts for the sum over spins.
Assuming isotropic collisions, the mean velocity of electrons emitted
from collisions equals the velocity of the lattice,
\begin{equation}
2\int \frac{\td{\bf k}}{ (2\pi)^3}{\bar f}{\bf v}=n\dot{\bf u}.
\label{relaxvel}
\end{equation}

The quasiparticle energy in the lattice moving with velocity
$\dot{\bf u}$ is
\begin{equation}
\varepsilon=\frac{|{\bf k}-e{\bf A}|^2}{ 2\emas }+
\chi\frac{|{\bf k}-e{\bf A}-\emas \dot{\bf u}|^2}{ 2\emas },
\label{quasienergy}
\end{equation}
where $\chi$ measures the renormalization of the inverse mass
$1/m=(1+\chi)/\emas $ in the normal state and the term proportional
to it describes the normal entrainment. The corresponding quasiparticle
velocity is
\begin{equation}
{\bf v}=\frac{\del\varepsilon}{\del{\bf k}}=
{{\bf k}-e{\bf A}-\chi m\dot{\bf u}\over m}.
\label{quasivelocity}
\end{equation}

The local equilibrium distribution $\bar f$ is centered around
the mean momentum ${\bar{\bf k}}$,
\begin{equation}
2\int{\td{\bf k}\over (2\pi)^3}{\bar f}~{\bf k}=n{\bar{\bf k}}.
\label{relaxmom}
\end{equation}
The condition \eqref{relaxvel} then gives
\begin{equation}
{\bar{\bf k}}=e{\bf A}+\emas \dot{\bf u},
\label{relaxvel2}
\end{equation}
where we have used $(1+\chi)m=\emas $.
The local equilibrium is thus given by the Fermi-Dirac distribution
\begin{equation}
{\bar f}({\bf k},{\bf r},t)=f_{FD}(\bar\varepsilon)
\label{relaxdis}
\end{equation}
with energy $\bar\varepsilon=|{\bf k}-{\bar{\bf k}}|^2/2m$ or
\begin{equation}
\bar\varepsilon({\bf k},{\bf r},t)={|{\bf k}-e{\bf A}({\bf r},t)-
\emas \dot{\bf u}({\bf r},t)|^2\over 2m}.
\label{relaxen}
\end{equation}

The total current is the sum of the ionic current $-en\dot{\bf u}$ and
the electronic current
\begin{equation}
{\bf j}=-en\dot{\bf u}+2e\int{\td{\bf k}\over (2\pi)^3}f\,{\bf v}.
\label{curdef}
\end{equation}
According to \eqref{relaxvel}
\begin{equation}
2e\int{\td{\bf k}\over (2\pi)^3}{\bar f}~{\bf v}=en\dot{\bf u}
\label{relaxcur}
\end{equation}
which exactly cancels the ionic current. The total current
due to the deviation from local equilibrium is thus
\begin{equation}
{\bf j}=2e\int{\td{\bf k}\over (2\pi)^3}{\delta f}\,{\bf v}.
\label{cur}
\end{equation}

The distribution $\delta f$ we will find from the Boltzmann
equation \eqref{Boltzmann}
\begin{equation}
\left({1\over \tau}+
{\del\over\del t}+{\bf v}\cdot\nabla-
\nabla\varepsilon\cdot{\del\over\del {\bf k}}\right)\delta f
=-\bar I
\label{Boltzmannexp}
\end{equation}
with the source term
\begin{equation}
\bar I
={\del\bar f\over\del t}+{\bf v}\cdot\nabla\bar f-
{\del\bar f\over\del {\bf k}}\cdot\nabla\varepsilon.
\label{Boltzmannexpsource}
\end{equation}
The local equilibrium depends on the time and space only via the
central momentum $\bar{\bf k}$, therefore
\begin{eqnarray}
{\del\bar f\over\del t}&=&{\del\bar f\over\del k_i}
\left(-e{\del A_i\over\del t}-\emas {\del\dot u_i\over\del t}
\right),
\label{Boltzmannexp1}\\
{\bf v}\cdot\nabla\bar f&=&{\del\bar f\over\del k_i}
\left(-ev_j\nabla_j A_i-\emas v_j\nabla_j\dot u_i\right),
\label{Boltzmannexp2}\\
-{\del\bar f\over\del {\bf k}}\cdot\nabla\varepsilon&=&
{\del\bar f\over\del k_i}
\left(ev_j\nabla_i A_j+(\emas -m)v_j'\nabla_i\dot u_j\right).
\nonumber\\
\label{Boltzmannexp3}
\end{eqnarray}
We have used the velocity relative to the lattice
\begin{equation}
{\bf v}'={\del\bar\epsilon\over\del{\bf k}}=
{\bf v}-\dot{\bf u}.
\label{relvel}
\end{equation}

Using relations $-\del_t{\bf A}=\bf E$, 
$v_j\nabla_i A_j-v_j\nabla_j A_i=[{\bf v}\times{\bf B}]_i$
and $\del_t\dot{\bf u}+\dot{\bf u}\cdot\nabla\dot{\bf u}=
\ddot{\bf u}$, the source term can be expressed as
\begin{eqnarray}
\bar I&=&{\del\bar f\over\del{\bf k}}
\biggl(e{\bf E}+e{\bf v}\times{\bf B}-\emas \ddot{\bf u}+
\emas {\bf v}'\times[\curl\dot{\bf u}]\biggr)
\nonumber\\
&-&{\del\bar f\over\del k_i}m v_j'\nabla_i\dot u_j.
\label{Boltzmannexp4}
\end{eqnarray}
As the local equilibrium distribution $\bar f$ depends only
on $\bar\varepsilon$, the source term can be further simplified
\begin{equation}
\bar I={\del\bar f\over\del\bar\varepsilon}{\bf v}'\cdot
\biggl(e{\bf E}+e\dot{\bf u}\times{\bf B}-\emas \ddot{\bf u}-
m({\bf v}'\cdot\nabla)\dot{\bf u}\biggr),
\label{Boltzmannexp6}
\end{equation}
where we have used orthogonality ${\bf v}'\cdot({\bf v}'\times{\bf B})=0$
and ${\bf v}'\cdot\left({\bf v}'\times[\curl\dot{\bf u}]\right)=0$.

The current \eqref{cur} in terms of the relative velocity
\eqref{relvel} is
\begin{equation}
{\bf j}=2e\int{\td{\bf k}\over (2\pi)^3}\delta f{\bf v}'.
\label{curdev}
\end{equation}
The term proportional to $\dot{\bf u}$ equals zero, because from
\eqref{relaxn} follows
$\int d{\bf k}\,\delta f=0$.

To evaluate the deviation to terms linear in $\dot{\bf u}$
we can neglect nonlinear terms in the left hand side of
\eqref{Boltzmannexp}
\begin{equation}
\left({1\over \tau}+i\omega+{\bf v}'\cdot\nabla-
\nabla\bar\varepsilon\cdot{\del\over\del {\bf k}}\right)
\delta f=-\bar I.
\label{Boltzmannlin}
\end{equation}
The distribution $\delta f$ depends on ${\bf r}$ and ${\bf k}$ in two
ways, via $\bar\varepsilon$ in $\bar f$, and via vectors ${\bf v}'$
and $\nabla\bar\varepsilon$. Dependence on $\bar\varepsilon$ can be
eliminated. Let us write the derivatives as
\begin{eqnarray}
\nabla\delta f&=&
{\del\delta f\over\del\bar\varepsilon}\nabla\bar\varepsilon+
\left({\del\delta f\over\del{\bf r}}\right)_{\bar\varepsilon},
\label{nablasplit}\\
{\del\delta f\over\del{\bf k}}&=&
{\del\delta f\over\del\bar\varepsilon}{\bf v}'+
\left({\del\delta f\over\del{\bf k}}\right)_{\bar\varepsilon}.
\label{derksplit}
\end{eqnarray}
The energy derivative cancels, therefore
\begin{equation}
\left({1\over \tau}+i\omega\right)\delta f
+{\bf v}'\cdot
\left({\del\delta f\over\del{\bf r}}\right)_{\bar\varepsilon}-
\nabla\bar\varepsilon\cdot
\left({\del\delta f\over\del{\bf k}}\right)_{\bar\varepsilon}=-
\bar I.
\label{Boltzmannlinsplit}
\end{equation}

We will expand the solution in small $\tau/(1+i\tau\omega)$. The
first order is
\begin{equation}
\delta f_1=-{\tau\bar I\over 1+i\tau\omega},
\label{Boltzmannlin1}
\end{equation}
and the second order is
\begin{eqnarray}
\delta f_2&=&-{\tau\over 1+i\tau\omega}\left({\bf v}'\cdot
\left({\del\delta f_1\over\del{\bf r}}\right)_{\bar\varepsilon}-
\nabla\bar\varepsilon\cdot
\left({\del\delta f_1\over\del{\bf k}}\right)_{\bar\varepsilon}\right)
\nonumber\\
&=&\left({\tau\over 1+i\tau\omega}\right)^2\left({\bf v}'\cdot
\left({\del\bar I\over\del{\bf r}}\right)_{\bar\varepsilon}-
\nabla\bar\varepsilon\cdot
\left({\del\bar I\over\del{\bf k}}\right)_{\bar\varepsilon}\right)
\nonumber\\
&=&\left({\tau\over 1+i\tau\omega}\right)^2{\del\bar f\over\del\bar\varepsilon}
v'_j\biggl(\nabla_j\bigl(v'_i F'_i-mv'_iv'_k\nabla_k\dot u_i\bigr)
\nonumber\\
&&+
e(\nabla_l A_j)
{\del\over\del k_l}\bigl(v'_i F'_i-mv'_iv'_k\nabla_k\dot u_i\bigr)\biggr)
\nonumber\\
\label{Boltzmannlin2}
\end{eqnarray}
with the force
\begin{equation}
{\bf F}'=e{\bf E}+e\dot{\bf u}\times{\bf B}-\emas \ddot{\bf u}.
\label{efforce}
\end{equation}

In the linear response we can neglect $\dot{\bf u}$ in derivatives,
$\nabla_jv'_i=-(e/m)\nabla_jA_i$ and $(\del v'_i/\del k_j)=
(1/m)\delta_{ij}$, therefore
\begin{eqnarray}
\delta f_2
&=&\left({\tau\over 1+i\tau\omega}\right)^2{\del\bar f\over\del\bar\varepsilon}
\biggl({e\over m}v'_j F'_i(\nabla_i A_j-\nabla_jA_i)
\nonumber\\
&&~~~~~~-ev'_jv'_k(\nabla_i A_j-\nabla_jA_i)(\nabla_k\dot u_i+\nabla_i\dot u_k)
\nonumber\\
&&~~~~~~-
mv'_jv'_iv'_k\nabla_j\nabla_k\dot u_i+v'_jv'_i\nabla_j F'_i
\biggr)
\nonumber\\
&=&\left({\tau\over 1+i\tau\omega}\right)^2{\del\bar f\over\del\bar\varepsilon}
\biggl({e\over m}{\bf v}'\cdot[{\bf B}\times{\bf F}']
\nonumber\\
&&+m({\bf v}'\cdot{\bf q})^2({\bf v}'\cdot\dot{\bf u})-i({\bf v}'\cdot{\bf q})({\bf v}'\cdot{\bf F}')
\nonumber\\
&&+ie{\bf B}\cdot\Bigl([\dot{\bf u}\times{\bf v}']({\bf v}'\cdot{\bf q})+
[{\bf q}\times{\bf v}']({\bf v}'\cdot\dot{\bf u})\Bigr)\biggr).
\nonumber\\
\label{Boltzmannlin2a}
\end{eqnarray}

The function $\delta f=\delta f_1+\delta f_2$ includes terms odd and even
in the velocity ${\bf v}'$. We keep only the odd terms which contribute to
the current,
\begin{eqnarray}
 &&\delta f_{odd}=-{\tau\over 1+i\tau\omega}
		{\del\bar f\over\del\bar\varepsilon}
	 	{\bf v}'\cdot{\bf F}'	\nonumber\\
&&+\left({\tau\over 1+i\tau\omega}\right)^2{\del\bar f\over\del\bar\varepsilon}
\biggl({e\over m}{\bf v}'\cdot[{\bf B}\times{\bf F}']+m({\bf v}'\cdot{\bf q})^2({\bf v}'\cdot\dot{\bf u})\biggr).
\nonumber\\
\label{Boltzmannlinodd}
\end{eqnarray}
The electric current is thus
\begin{equation}
{\bf j}={\sigma_{n}\over e}\left({\bf F}'-{\tau\over 1+i\tau\omega}{e\over m}{\bf B}\times{\bf F}'\right)-\nu\dot{\bf u},
\label{curdevfin}
\end{equation}
where
\begin{equation}
\sigma_{n}=-{2\tau e^2\over 1+i\tau\omega}{1\over 3}
\int{\td{\bf k}\over (2\pi)^3}
{\del\bar f\over\del\bar\varepsilon}v'^2=
{\tau e^2n\over m(1+i\tau\omega)}
\label{sigma}
\end{equation}
is the usual conductivity in the absence of the magnetic field.
The Hall component is implied by the force term ${\bf B}\times{\bf F}'$.

The last term in \eqref{curdevfin} results from inhomogeneous velocity
of the lattice, namely impurities and phonons. Its coefficient reminds the
shear viscosity
\begin{eqnarray}
\nu&=&-{2\tau^2 emq^2\over (1+i\tau\omega)^2}\int{\td{\bf k}\over (2\pi)^3}
{\del\bar f\over\del\bar\varepsilon}v_x^{\prime 2}v_z^{\prime 2}.
\label{shareviscgen}
\end{eqnarray}
The integral over velocities in \eqref{shareviscgen} in the zero temperature
limit is
\begin{eqnarray}
&&\!\!\!\!\!\!\!\!\!\!\!-2\int{\td{\bf k}\over (2\pi)^3}
{\del\bar f\over\del\bar\varepsilon}v_x^{\prime 2}v_z^{\prime 2}
\nonumber\\
&=&
{2\over (2\pi)^3}\!\int\limits_{-1}^1\!dzz^2(1\!-\!z^2)\!\int\limits_{-\pi}^\pi\! d\varphi \sin^2\varphi
\nonumber\\
&&\times
\int\limits_0^\infty\! d\bar k \delta(\bar\varepsilon-E_{F}){\bar k^6\over m^4}
\nonumber\\
&=&{1\over 15\pi^2}{k^5_{F}\over m^3}
\nonumber\\
&=&{nv^2_{F}\over 5m},
\label{muint}
\end{eqnarray}
where we have used the density $n=k_{F}^3/(3\pi^2)$ and the
Fermi velocity $v_{F}=k_{F}/m$. Finally, we express the
shear coefficient in terms of the mean free path
$l=\tau v_{F}$
\begin{eqnarray}
\nu&=&{eq^2nl^2\over 5(1+i\tau\omega)^2}.
\label{sharevisc}
\end{eqnarray}
For short lifetime $\tau\omega\to 0$, the coefficient $\nu$ agrees with
the result of Fil \cite{Fil01}.

%-----------------------------------------------------------------
\section{Chemical potential}
\label{Chemical}
%-----------------------------------------------------------------

Here we show that the chemical potential can be excluded from
assumptions dealing with the fields averaged over elementary
cells of the Abrikosov vortex lattice.

Let us split the chemical potential as $\mu=\mu_{\bf j}+\mu_{\bf u}$,
where the first term has the form standard in the TDGL theory
\begin{equation}
\nabla^2\mu_{\bf j}={e\over\sigma_{n}}\nabla\cdot\js
 \label{potentialj}
\end{equation}
and the second term appears only in moving crystals and represents
a change of the chemical potential due to the Lorentz force
\begin{equation}
\nabla^2\mu_{\bf u}=e\nabla\cdot[\dot{\bf u}\times{\bf B}].
 \label{efmusub}
\end{equation}
Both potentials need boundary conditions which specify constant and linear
terms. We use zero mean values, $\langle\mu_{\bf j}\rangle_s=0$ and
$\langle\mu_{\bf u}\rangle_s=0$, where brackets denote average over sample
volume. Since the system is periodic on the Abrikosov vortex lattice,
this averaging is identical to averaging over single elementary cell
and implies zero mean gradients $\langle\nabla\mu_{\bf j}\rangle={\bf 0}$
and $\langle\nabla\mu_{\bf u}\rangle={\bf 0}$.

It is necessary to show that the conditions $\langle\nabla\mu_{\bf j}\rangle={\bf 0}$
and $\langle\nabla\mu_{\bf u}\rangle={\bf 0}$ are not in conflict with
equations \eqref{potentialj} and \eqref{efmusub} respectively.
The source term in the right hand side of \eqref{potentialj}
is a sum of the transport supercurrent $\langle\js\rangle$ and the 
circulating current due to the Abrikosov vortex lattice. In the 
homogeneous Abrikosov lattice the transport supercurrent has zero divergence
$\nabla\cdot\langle\js\rangle=0$ because of the translation invariance.
The circulating component has zero divergence in the approximation
of rigidly moving Abrikosov lattice. Beyond this approximation one
finds contributions that are nonzero but periodic on the Abrikosov
lattice giving the zero mean value, $\langle\nabla\cdot\js
\rangle=0$. Zero mean value of the source term in \eqref{potentialj}
is not in conflict with the boundary condition $\langle\mu_{\bf j}\rangle=0$.

The source term in the right hand side of the equation \eqref{efmusub} is
rather complex. It simplifies in the linear approximation in $\bf u$ as
\begin{eqnarray}
\nabla\cdot[\dot{\bf u}\times{\bf B}]&=&-
\dot{\bf u}\cdot[\curl{\bf B}]+{\bf B}\cdot[\curl\dot{\bf u}]
\nonumber\\
&=&-\mu_0\dot{\bf u}\cdot\left(\js+\jn+\jl\right)
\nonumber\\
&\approx&-\mu_0\dot{\bf u}\cdot\js^0,
 \label{uB}
\end{eqnarray}
where $\js^0$ is the supercurrent in the static Abrikosov lattice.
We have used that the wave propagates along the magnetic field ${\bf B}\|{\bf q}$,
therefore ${\bf B}\cdot[\curl\dot{\bf u}]=0$. In the last step
we have neglected terms beyond the linear response.
Since there is no transport current in the static Abrikosov
lattice $\langle\js^0\rangle={\bf 0}$, the source term
has zero mean value, $\langle\dot{\bf u}\cdot\js^0\rangle=
\dot{\bf u}\cdot\langle\js^0\rangle=0$. The boundary
condition $\langle\mu_{\bf u}\rangle=0$ is thus not in conflict with
the source term.

%-----------------------------------------------------------------
\section{Parameters of niobium}
\label{Filparam}
%-----------------------------------------------------------------

First we list characteristic values. At the upper critical field
at zero temperature $B=B_{c2}=.49$~T, the cyclotron frequency
$\omega_{c}=eB/m$ is $\omega_{c}=7.13\cdot 10^{10}$~s$^{-1}$,
with the effective mass of niobium  $m=1.2~\emas $. We note
that niobium has a complicated Fermi surface appearing in the
first, second and third Brillouin zone so that different
effective masses appear, e.g. $m=3.2~\emas $ and $m=1.7~\emas $
from de Haas-van Alphen effect with and without phonon dressing
\cite{KS70}.
In all cases the cyclotron frequency is much higher than the
frequency of applied sound $\omega=2\pi~5.5\cdot 10^7$~s$^{-1}$.
The velocity of the transverse sound in niobium is $v_{s}=2100$~m/s.
This gives the wave vector $q=\omega/v_{s}=1.6\cdot 10^5$/m and
wave length $2\pi/q=3.8\cdot 10^{-5}$~m.

Because of the complicated energy band structure, it is preferable
to use characteristics of the Fermi surface rather than the effective
mass and electron density. The single-spin density of states is
$N_0=5.7\cdot 10^{47}$/Jm$^3$, and the average of the Fermi velocity over
the niobium Fermi surface is $v_{F}=0.59\cdot 10^6$~m/s, see Weber {\em et al}
\cite{WSLSPJE91}. They enter the conductivity as $\sigma_{n}=
\frac23 e^2N_0v_{F}^2\tau$.

The relaxation time $\tau$ depends on impurities. The niobium sample
measured by Fil {\em et al} \cite{Fil06} reveals a step of
acousto-electric effect in the zero magnetic field. Going from
the superconducting state to the normal one, the magnitude reduces
by 10\% and the phase increases by 7$^\circ$. Within the
present theory it is reproduced by $\tau=1.2\cdot 10^{-13}$~s,
which corresponds to the residual resistivity ratio $RRR=62$.
This short relaxation time leads to rather small dimensionless numbers
$\tau\omega_{c}=8.6\cdot 10^{-3}$ and $\tau\omega=4.1\cdot 10^{-5}$.

Different values one finds for pure samples.  Weber {\em et al}
\cite{WSLSPJE91} measured a sample of the residual resistivity
ratio $RRR=2080$, giving the low temperature conductivity
$\sigma_{n}=RRR/\rho_{n}=3\cdot 10^{13}/\Omega$m.
This high conductivity corresponds to the relaxation time
$\tau=8.9\cdot 10^{-9}$~s with the dimensionless number $\tau\omega=3.1$.
The mean number of circulations between collisions is
$\tau\omega_{c}=635$. Since the relaxation time $\tau$ is
proportional to the $RRR$, it is possible to prepare samples
with $\tau$ from $10^{-8}$~s to $10^{-14}$~s. Moreover, for
a thin sample the magnetic field can be weak so that dimensionless
numbers can have general values from small to values over unity.

The mean free path $l=v_{F}\tau$ spreads from $6\cdot 10^{-9}$~m
to $6\cdot 10^{-3}$~m. For the mean free path exceeding the wave length
the theory of the normal acousto-electric effect is not fully
justified because it is based on local approach with the lowest
order nonlocal correction $\nu$. To stay in the region of
validity we assume $\tau\ll 10^{-11}$~s for which $l\ll 1/q$.
The value RRR=620 used for demonstration corresponds to
$\tau=1.2\cdot 10^{-12}$~s giving small dimensionless numbers
$\tau\omega_{c}=8.6\cdot 10^{-2}$ and $\tau\omega=4.1\cdot 10^{-4}$.

Let us identify parameters for the superconducting state.
For the dirty sample of Fil {\em et al}, the critical fields
correspond to the GL parameter $\kappa=1.5$ given by the
coherence length $\xi_0=2.6\cdot 10^{-8}$~m and the London penetration
depth $\lambda=3.9\cdot 10^{-8}$~m. Here $\lambda=\sqrt{m^*/(2\mu_0e^2\ns)}$
with the Cooperon mass $m^*=2m\hbar v_{F}/(\pi\Delta_0 l_{free})=
5.52~\emas $. We have used the BCS gap $\Delta_0=1.76k_{B}T_{c}$.
In the superconducting regime far from the critical temperature nonlocal
contributions are negligible because $q\lambda=6.13\cdot 10^{-3}$. 

Going to the clean limit there will be no dramatic changes. The Cooperon
mass $m^*$ reaches the value of $2m$. The London penetration depth
decreases to $\lambda=2.3\cdot 10^{-8}$~m and the GL parameter reduces close
to the limiting value $\kappa\sim 1/\sqrt{2}$. Samples of width
comparable to the wave length $\sim 10^{-5}$~m, but large in area
$\sim 1$~cm$^2$, are penetrated by the magnetic field
either in the form of Abrikosov vortices or in the form
of slabs. We discuss the case in the vicinity of the critical temperature 
where sample becomes effectively thin as $\lambda$ is large so that vortices 
become preferable.

%-----------------------------------------------------------------
\section{Forces on vortex}
\label{Foronvor}
%-----------------------------------------------------------------

We take the friction according to Kopnin \cite{Kop01} (formula
12.38 with $B\to B_{c2}$ limit of $\sigma_f$ given by
formula 12.35) $\eta=\eta_{lat}=\sigma_{n}B_{c2}=
1.8\cdot 10^6$~C/m$^3$. The quasiparticle friction is neglected
in this approximation.

The coefficient of the Magnus-like force $\alpha_{M}=
e\ns+\alpha_{I}=3.5\cdot 10^9$~C/m$^3$ is dominated
by the Iordanskii term $\alpha_{I}=e(n-\ns)$, see
\cite{Kop01} formula below (14.97).
We neglect the Kopnin-Kravtsov force.

The presented numerical results have been obtained with rather
small Labusch coefficient
$\alpha_{L}\ll e\ns/\omega$. We have found that
acousto-electric effect remains the same within accuracy of
figures even for values as large as $\alpha_{L}\sim 10^3
e\ns/\omega$.

\end{document}